# Doppler velocimetry of spin propagation in a two-dimensional electron gas

Luyi Yang[1,2], J. D. Koralek[2], and J. Orenstein[1,2,*], D. R. Tibbetts[3], J. L. Reno[3] and M. P. Lilly[3]

[1]*Department of Physics, University of California, Berkeley, California 94720, USA.*
[2]*Materials Science Division, Lawrence Berkeley National Laboratory, Berkeley, California 94720, USA.*
[3]*Sandia National Laboratories, Albuquerque, New Mexico 87123, USA.*

[*]*email:* JWOrenstein@lbl.gov

**Controlling the flow of electrons by manipulation of their spin is a key to the development of spin-based electronics. While recent demonstrations of electrical-gate control in spin-transistor configurations show great promise, operation at room temperature remains elusive. Further progress requires a deeper understanding of the propagation of spin polarization, particularly in the high mobility semiconductors used for devices. Here we report the application of Doppler velocimetry to resolve the motion of spin-polarized electrons in GaAs quantum wells driven by a drifting Fermi sea. We find that the spin mobility tracks the high electron mobility precisely as a function of *T*. However, we also observe that the coherent precession of spins driven by spin-orbit interaction, which is essential for the operation of a broad class of spin logic devices, breaks down at temperatures above 150 K for reasons that are not understood theoretically.**

The transistor, the iconic invention of 20[th] century science, is a semiconductor device in which the flow of electrons is modulated by voltages applied *via* electrodes known appropriately as gates. In a conventional transistor the gate electrode controls the number of mobile electrons in the current carrying pathway, or "channel." In pursuit of transistors with faster response and lower rates of energy dissipation, there has been intense investigation aimed at modulating current through manipulation of spin by applied electric fields [1,2], a coupling that occurs because of the spin-orbit (SO) interaction. Recently, gate-controlled modulation of current *via* SO coupling has been demonstrated in prototype device structures that operate below room temperature [3,4].

Further progress towards spintronic logic requires a deeper understanding of the basic physical principles upon which such devices are based. Essentially the question is this: how far, and how fast, can spin polarization propagate in a current-carrying electron gas? This question was first addressed in pioneering work that used magneto-optic imaging to follow the drift of spin polarization packets in real space [5]. These experiments were enabled by the enhanced spin lifetimes (in excess of 10 ns) that arise near the metal-insulator transition of a doped semiconductor at the expense of electron mobility, $\mu_e$. However, the high $\mu_e$ electron gas needed for fast devices is in a very different dynamical regime, where spin lifetimes are ~ 10-100 ps, during which time spin may propagate only 10-100 nm (depending on the temperature, *T*, and applied field, *E*). To resolve spin propagation on picosecond time and nanometer length scales we have developed a technique to measure velocity *via* the Doppler shift of light scattered from propagating waves of spin density. Our method extends transient grating spectroscopy (TGS) [6], which has traditionally been used to measure rates of diffusion, to the measurement of drift velocity.

**Transient grating spectroscopy and Doppler velocimetry**

Excitation of a semiconductor with a single beam of above band-gap energy photons injects an equal population of electrons and holes, whose spatial distribution follows the intensity of the laser spot. While recombination rates can be readily determined from the lifetime of this excited state, information about the motion of electrons and holes is inferred only indirectly. TGS adds the element of spatial resolution to time-resolved optical measurements, enabling direct probing of transport. In TGS two non-collinear beams of light pulses interfere at the sample surface, creating a pattern of intensity and photon helicity that depends on the relative angle and polarization state of the two beams. When the two beams are polarized parallel to each other, interference creates a standing wave of laser intensity, generating the sinusoidal pattern of photoinduced electron-hole (*e-h*) density shown in Fig. 1a. On the other hand, orthogonal polarization generates a standing wave of photon helicity, while maintaining spatially uniform intensity (on the scale of the laser spot) and therefore the *e-h* density. Optical selection rules in GaAs cause photon helicity to be imprinted in the out of plane (*z*) component of the angular momentum of the photoinduced *e-h* gas [7]. Because hole spins depolarize on a sub-picosecond time scale, the excited state for $t > 1$ ps comprises an electron spin density wave (SDW) accompanied by a charge compensating gas of unpolarized holes, as illustrated in Fig. 1b. The SDW induces variation in the local index of refraction and therefore acts as a transient optical grating, whose subsequent dynamics can be monitored *via* the diffraction of a time-delayed probe pulse.

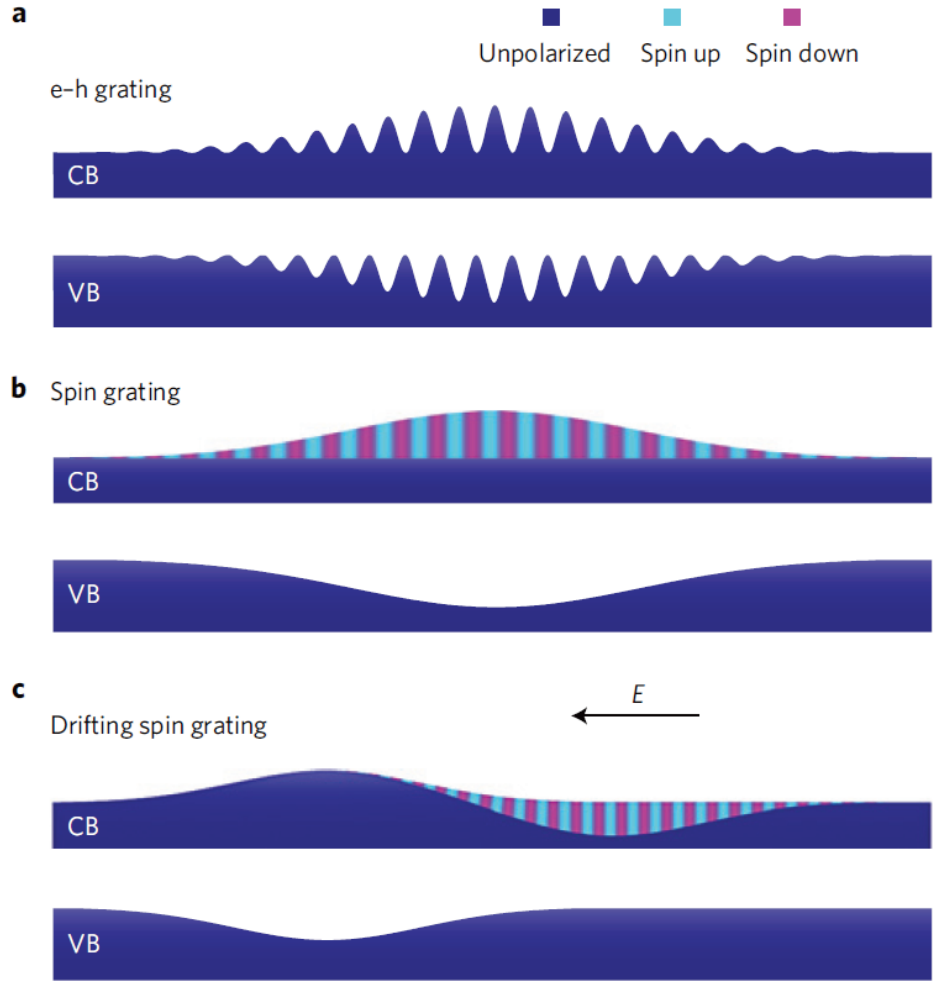

**Figure 1 | Illustration of photoinduced transient gratings in a doped quantum well. a**, The photoinduced *e-h* grating consists of a sinusoidal variation of the *e-h* density modulated by the Gaussian envelope of the laser spot. In the presence of an applied *E* field, the *e-h* grating moves at the ambipolar velocity. **b**, The photoinduced spin grating consists of a sinusoidal modulation of the out-of-plane component of spin density coexisting with the Gaussian distribution of electrons and holes. The hole spins randomize on the sub-picosecond timescale and so are shown as unpolarized. Under the influence of an applied *E* field, the spin grating and the Gaussian *e-h* packet move in the direction of the Fermi sea of electrons. While the spin grating moves at or near the velocity of the Fermi sea, the *e-h* packet moves at the much slower ambipolar velocity. This leads to an increasing spatial separation of the spin and charge degrees of freedom as shown in **c**. CB: conduction band; VB: valence band.

In the experiments reported here we photo-inject the SDW into a 2DEG subject to an in-plane $E$ field that is parallel to the grating wavevector and measure the resulting propagation of spin polarization. If the polarization wave undergoes normal drift and diffusion, the spin density will evolve according to $\mathbf{S}(x,t) = \mathbf{S}_0 \exp[-t/\tau(q)]\cos\{q[x-x_0(t)]\}$, where $q\hat{\mathbf{x}}$ is the wavevector and $x_0(t)$ is the displacement. Measurement of the amplitude of a diffracted pulse yields the wavevector dependent lifetime $\tau(q)$, from which the spin memory time and diffusion coefficient can be determined. Information about $x_0(t)$ is contained in the *phase*, rather than the amplitude, of the diffracted light. For example, light diffracted from an SDW drifting at constant velocity, $x_0(t) = v_d t$, will contain the optical phase factor $\phi(t) = qv_d t$. The linear advance of phase with time is equivalent to a Doppler shift, $\Delta\omega = v_d q$, as illustrated in Fig. 2. To measure the phase of the diffracted light we used a heterodyne technique, in which diffracted pulses are mixed in a Si photodiode with a beam of transmitted pulses acting as a local oscillator [8,9,10]. The 3 mrad phase noise level of our detection system corresponds to ~1 nm resolution of the position of the SDW.

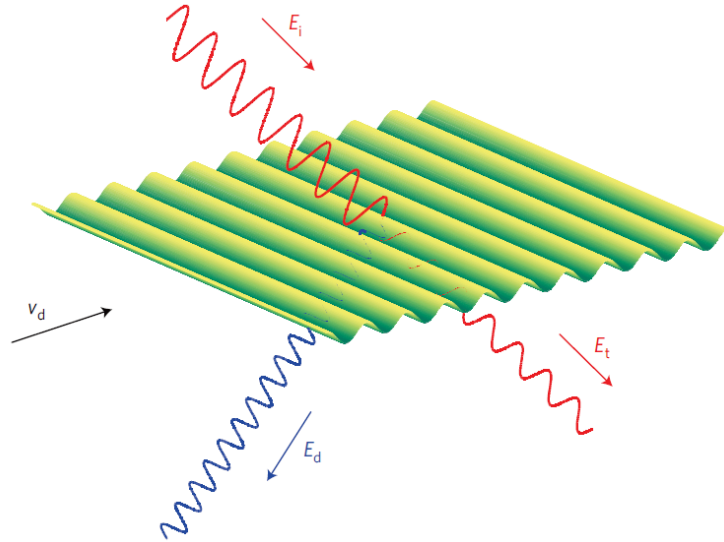

**Figure 2 | Doppler velocimetry.** The illustration depicts a transient grating moving with velocity $v_d$ parallel to its wavevector $\mathbf{q}$. A probing light field $E_i(\omega)$ is incident on the grating. Upon interaction with the moving grating the probe is divided between a transmitted beam at the same optical frequency $E_t(\omega)$, and a Doppler shifted diffracted beam, $E_d(\omega + \mathbf{q}\cdot\mathbf{v}_d t)$. From measurements of the Doppler shift we obtain the grating drift velocity $v_d$.

## Spin diffusion

Our measurements were performed on samples containing a single 9 nm wide electron-doped GaAs/GaAlAs quantum well. The carrier density and mobility of the 2DEG were $1.9 \times 10^{11}$ cm$^{-2}$ and $5.5 \times 10^5$ cm$^2$/V-s at 5 K, respectively. The sample structure and processing are described in detail in the Methods section. Shown in Fig. 3a is $A(q,t)$, the amplitude of the probe diffracted from a photoinjected SDW with wavevector parallel to the [110] crystal axis, as measured at $T = 30$ K. The time dependence of the amplitude is the sum of two exponentially decaying components, $A(q,t) = A_+\exp[-t/\tau_+(q)] + A_-\exp[-t/\tau_-(q)]$, with nearly equal weighting factors ($A_+ \approx A_-$). The two lifetimes, $\tau_\pm(q)$, are plotted as a function of $q$ in Fig. 3b. The decay rate of the shorter-lived component $1/\tau_+(q)$ is proportional to $q^2$ as expected for a simple diffusive process. However, the $q$-dependence of $1/\tau_-(q)$ is anomalous, with a minimum rate found at a nonzero wavevector, $q_0 \approx 0.6 \times 10^4$ cm$^{-1}$.

The existence of two rates is a consequence of SO coupling, which in a GaAs QW has the form of an effective magnetic field that induces spin precession at a rate that depends on the electron's momentum, $\boldsymbol{p}$. In a symmetric QW the Dresselhaus SO coupling [11] dominates, which is characterized by the precession rate vector field, $\boldsymbol{\Omega}(\boldsymbol{p}) = 2\hbar^{-2}\beta_1(p_y\hat{\boldsymbol{x}} + p_x\hat{\boldsymbol{y}})$, where $\beta_1$ is the linear Dresselhaus coupling strength and $\hat{\boldsymbol{x}}$ and $\hat{\boldsymbol{y}}$ are the [110] and [1$\bar{1}$0] crystal axes, respectively. The connection between the precession vector and momentum induces a strong correlation between the diffusion of electrons in real space and of the orientation of their spins on the Bloch sphere. Theoretical analysis of this correlation yields a pair of normal modes at each $q$ that are helical waves of spin density with opposite sense of rotation [12,13,14,15,16,17]. The lifetime of the helix whose sense of rotation matches that of the electron's precession is strongly enhanced by SO coupling while the lifetime of the helix with opposite rotation is reduced. Both lifetimes, $\tau_\pm(q)$, are observed in the TSG experiment because the photogenerated initial state – a wave of pure $S_z$ – is a superposition of the two helices of opposite pitch. The solid lines through the data in Fig. 3b are fits to the spin helix theory [13,14,15] with $\beta_1 = 3.4 \times 10^{-3}$ eV Å.

From analysis of the measured $\tau_\pm(q)$ we also obtain the spin diffusion coefficient, $D_s(T)$, plotted in Fig. 3c. For comparison we plot the electron diffusion coefficient $D_e(T)$ obtained by applying the Einstein relation to the electron transport mobility $\mu_e$. $D_s(T)$ is smaller than $D_e(T)$ as a result of spin Coulomb drag (SCD) [18], which is a frictional force between oppositely oriented spins that is generated by electron-electron (*e-e*) collisions. As spin diffusion requires a counter flow of opposite spin populations, it is damped by SCD, whereas charge transport is protected from *e-e* collisions by momentum conservation. The reduction of spin relative to electron diffusion coefficient seen here is considerably larger than in previously reported measurements [19] – $D_s$ is only about 5% of $D_e$ when measured above the Fermi temperature of 80 K. The SCD effect is more pronounced in the cleaner sample studied here because, while its low $T$ resistivity is approximately eight times smaller than the previously studied QW with the same electron density, the intrinsic spin-drag transresistivity $\rho_{\uparrow\downarrow}(T)$ is unchanged. This is evident when we invert the measured $D_e/D_s$ to extract the transresistivity (See Supplementary Information Section I). The $\rho_{\uparrow\downarrow}(T)$ thus obtained (plotted in Fig. 3c inset) is quantitatively consistent with earlier reports and in excellent agreement with the random phase approximation (RPA) theory of SCD in two-dimensions in the range of $T$ below the Fermi temperature [20,21].

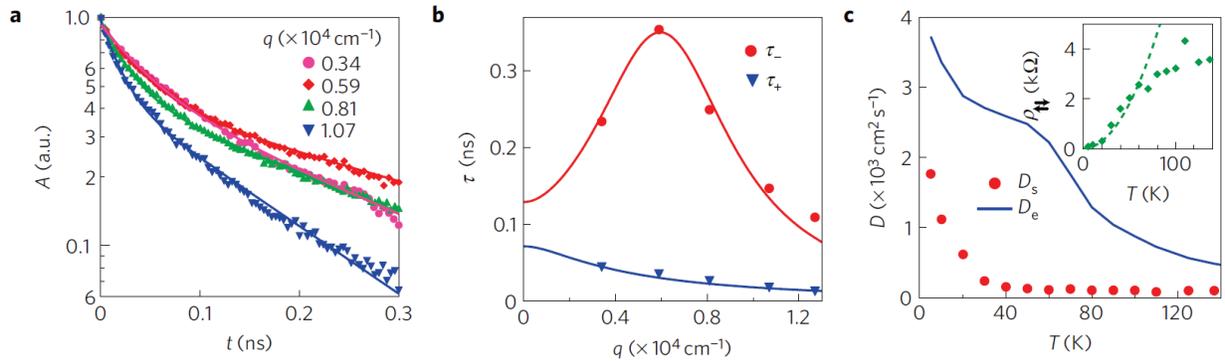

**Figure 3 | Spin diffusion and spin-Coulomb drag. a**, Decay of the amplitude of transient spin gratings measured at several values of the wavevector $q$. The solid lines are fits to a model of two exponentially decaying helical modes of equal amplitude. **b**, Lifetimes $\tau_\pm(q)$ for the spin helix modes of opposite pitch obtained from fits to the data in **a**. The short-lived mode, $\tau_+(q)$, is proportional to $1/q^2$ as expected for diffusion. The lifetime of the long-lived mode, $\tau_-(q)$, is peaked at a non-zero $q$ which is commensurate with the inverse spin-precession length in the SO field. **c**, Comparison of spin diffusion coefficient $D_s(T)$ determined from TGS with electron diffusion coefficient $D_e(T)$ determined by transport measurements. Inset: spin transresistivity $\rho_{\uparrow\downarrow}(T)$ extracted from ratio of $D_s(T)/D_e(T)$. The dashed line is the RPA prediction for $\rho_{\uparrow\downarrow}(T)$.

**Spin drift**

We turn now to Doppler shift measurements of spin helix drift under the influence of an $E$ field applied parallel to the SDW wavevector. Fig. 4a shows the phase, $\phi(q,t)$, of light diffracted from a transient spin grating as a function of $t$ for several values of $q$, measured at 30 K. For wavevectors larger than $q_0$, the phase increases linearly with time, indicating near uniform drift in the same direction as the Fermi sea of electrons. However, $\phi(q,t)$ is clearly more complex for $q < q_0$. While $\dot\phi$ starts out positive, it quickly crosses zero and becomes negative for $t \gtrsim 50$ ps, indicating counter-propagation with respect to the Fermi sea. It is natural to associate the anomalous behavior of the phase with the presence of the two helical modes discussed previously, and to describe the overall $\dot\phi(q,t)$ as the weighted average of their individual rates of phase advance, $\dot\phi_\pm(q)$,

$$\dot\phi(q,t) = \frac{A_+\exp[-t/\tau_+(q)]\dot\phi_+(q) + A_-\exp[-t/\tau_-(q)]\dot\phi_-(q)}{A_+\exp[-t/\tau_+(q)] + A_-\exp[-t/\tau_-(q)]}. \qquad (1)$$

The lines through the data in Fig. 4a are fits obtained with this expression, using the values of $\tau_\pm(q)$ obtained previously. The high quality of the fits suggests that the complicated behavior of $\dot\phi(q,t)$ reflects contributions from the two helices of opposite pitch, each propagating with its own uniform phase velocity. Shown in Fig. 4b are values $\dot\phi_\pm(q)$ obtained using Eq. 1 and the solid lines are fits to a theory of spin helix propagation described qualitatively below.

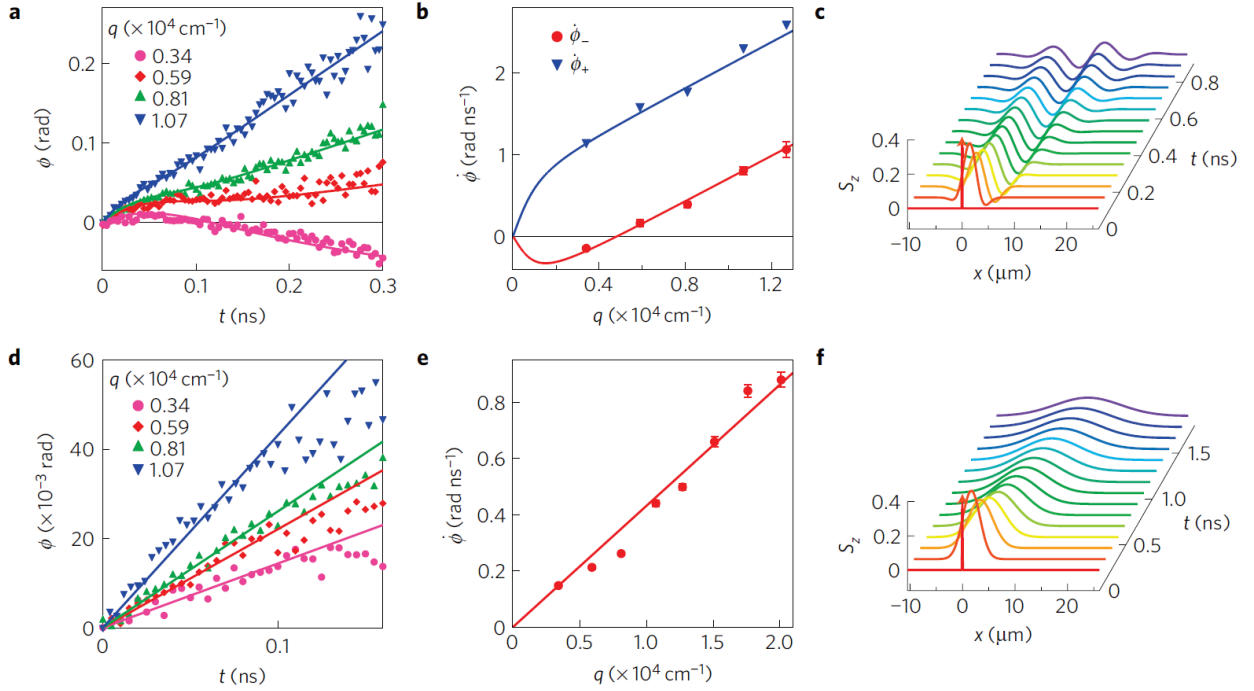

**Figure 4 | Spin drift in different temperature regimes. a**, Spin drift observed by Doppler velocimetry in the presence of an applied electric field ($E = 2$ V/cm) at $T=30$K. The quantity $\phi$ is the optical phase shift of the probe beam diffracted from the drifting spin grating. After the short-lived helix has decayed, the slope of $\dot\phi(t)$ is proportional to the velocity of the long-lived helix. The solid lines are a fit to Eq. 1 describing two independently propagating spin helix modes. The negative slope of the data at low $q$ demonstrates that the long-lived helix mode moves backwards (relative to the Fermi sea) for $q < q_0$. **b**, Wavevector dependence of the phase velocity associated with the two helical modes, $\dot\phi_\pm(q)$, obtained from the fits in **a**. The solid lines are a fit to the theoretical model of spin propagation in the presence of Dresselhaus SO coupling described in the text. **c**, Real-space spin propagator corresponding to the Fourier transform of the fit in **b**. The spin propagator, which describes the space-time evolution of spin polarization following $\delta$-function injection, has the form of an envelope function that moves at velocity $v_d$ while modulating a stationary helical SDW. Because this stationary pattern decays exponentially, it is necessary to multiply the propagator by $\exp[t/\tau_-(q_0)]$ in order to visualize the motion at long times. **d**, and **e**, show the same quantities as in **a**, and **b**, but measured at 150 K. **f**, Spin propagator at 150 K showing that transport takes place without coherent spin precession. Error bars (s.d.) in **b** and **e** represent the uncertainty in the fitting of **a** and **d**.

While for a Fermi sea at rest the average of $\mathbf{\Omega}(\mathbf{p})$ over occupied states is clearly zero, an electron gas drifting with velocity $v_d \hat{x}$ will experience a nonzero $\langle \mathbf{\Omega} \rangle = 2\beta_1 \hbar^{-2} m^* v_d \hat{y}$.

Consider first spin helices injected into a drifting Fermi sea in the absence of SO coupling. In this case the angle, $\theta$, of the local spin polarization with respect to $\hat{z}$ would be static in a frame moving with $\boldsymbol{v}_d$, $\theta(x') = \pm qx'$, where $x' = x - v_d t$. However, the nonzero SO coupling will cause the spins to precess as they drift, such that $\theta(x',t) = \pm qx' + \langle\Omega\rangle t$. When viewed in the stationary frame, $\theta(x,t) = \pm qx - v_d(q_0 \pm q)t$, where $q_0 \equiv 2\beta_1 \hbar^{-2} m^*$, which corresponds to the two rates of phase advance, $\dot{\phi}_\pm(q,t) = v_d(q \pm q_0)$. Thus for example, the long-lived helix will appear to be stationary when $q = q_0$ and counter-propagate for $q < q_0$. The lines through the data points in Fig. 4b are the predictions of quantitative theories of helix drift [22,23], which differ from the qualitative picture outlined above only at very low values of $q$ not accessible in our experiments.

The propagation of spin in the Dresselhaus field can also be visualized in the spatial rather than wavevector domain. As we have shown, TGS measures $S_z(q,t)$ over a broad range of $q$. The Fourier transform of $S_z(q,t)$ yields the real space spin propagator, $S_z(x,t)$, which describes the time evolution of spin polarization following pulsed injection of a narrow stripe of $z$-oriented spin density along the $y$-axis. Fourier transformation of the theoretical fits to the amplitude and phase of $S_z(q,t)$ shown in Figs. 3b and 4b yield the propagator illustrated in Fig. 4c., which has the form of an envelope function that moves with uniform velocity $v_d$ while modulating a stationary SDW (See Refs. 22,23 and Supplementary Information Section II). The polarization wave that emerges as the envelope propagates is closely related the stripe-like patterns imaged in steady state measurements on low-$\mu_e$ semiconductors [24].

A surprising feature of our results is that the spin propagation dynamics described above change drastically as $T$ is increased towards room temperature. As shown in Fig. 4d and 4e, at 150 K the two helices with phase dispersion $\phi(q,t) = v_d(q \pm q_0)t$ have been replaced by a single mode with $\phi(q,t) = v_d qt$. The latter corresponds to drift with velocity $v_d$ without spin precession. Fourier transformation of fits to $S_z(q,t)$ at 150 K yields a spin packet that propagates at $v_d$ and does not modulate a polarization wave, as shown in Fig. 4f (See Supplementary Section II). Thus a spin-transistor based on control of SO-induced precession will not operate in this $T$ regime. We note that the clear cross-over in spin dynamics that has taken place can be seen only in spatial or wavevector-resolved measurements, as the lifetime of the uniform spin polarization, $S_z(q = 0, t)$ increases monotonically from 30 to 150 K [25].

The absence of spin precession at 150 K cannot be attributed to a change in the SO coupling strength, $\beta_1$, which is an intrinsic property of the GaAs bandstructure. Instead, our results suggest that the effective precession vector $\langle \mathbf{\Omega} \rangle$ does not survive increased thermal averaging. One possible reason for this is the cubic (in $p$) Dresselhaus coupling, which causes the net precession angle between scattering events to depend on the electron's velocity, and has been shown to degrade the spin-spatial correlations described previously [17,26].

## Temperature dependence of spin mobility

We have seen that, when viewed in the spatial domain, an injected spin packet moves with $v_d = \partial \dot{\phi} / \partial q$, regardless of whether the propagation is accompanied by coherent spin precession. Thus at each $T$ we can determine a spin packet velocity from the dispersion of $\dot{\phi}(q)$, obtain a spin packet mobility, $\mu_s \equiv v_d / E$, and compare with the electron mobility, $\mu_e$ as determined from dc transport. In the course of such measurements we discovered that $\mu_s$ depends strongly on the intensity, $I$, of the laser pulse that generates the spin grating. Fig. 5a is a plot of $\mu_s$ as a function of $I$ for various $T$. As $I$ is reduced from its maximum value $I_0$ = 0.25 µJ/cm², $\mu_s$ initially increases and then approaches an asymptotic value $\mu_{s0}$ in the limit that $I \to 0$. The curves through the data points are fits to the relation, $\mu_s(I) = \mu_{s0}(T)/(1 + 2.86 I/I_0)$. The red circles in Fig. 5b represent $\mu_{s0}(T)$ as determined from fits to the intensity dependence. The plot shows that $\mu_{s0}(T) = \mu_e(T)$ the over the entire $T$ range of the experiment. Furthermore, this equality holds even as the nature of spin propagation crosses over from the precession regime, where $\dot{\phi}_\pm(q) = v_d(q \pm q_0)$, to the incoherent regime, where $\dot{\phi}(q) = v_d q$.

We argue below that the dependence of $\mu_s$ on $I$ indicates that the direct force of the electric field on the spin polarization is zero, and that spin waves (or packets) are propelled solely by momentum transfer from the surrounding Fermi sea. The basis of this claim is that the same dependence of mobility on pump intensity is observed when the Fermi sea drives another neutral excitation of the 2DEG, namely packets of *e-h* density [27,28]. Individual packets of electron and hole density cannot separate in weak applied fields – the constraint of local charge neutrality forces the two charge species drift together at a speed $\mu_a E$, where $\mu_a$ is the ambipolar mobility. Because the driving force of the Fermi sea scales with the equilibrium carrier concentration, $n_0$, while the packet's inertia varies as the photoinduced carrier concentration, $\Delta n$, the packet drift velocity depends on the ratio $\Delta n/n_0$, which is proportional to $I$. Solving the appropriate force balance equations yields $\mu_a \propto I^{-1}$ in the limit that $\Delta n \gg n_0$ and $\mu_a \to \mu_{a0}$ in the limit $\Delta n \ll n_0$, i.e., the same trends that we observe in the spin packet mobility. In Fig. 5b we compare $\mu_{a0}(T)$ determined from TGS measurements on the same QW using parallel polarization (see Fig. 1a) with the electron and spin mobility. The slower rate of ambipolar propagation reflects the fact that, in contrast with electron spin propagation, the low mobility holes must be dragged along with the drifting electrons.

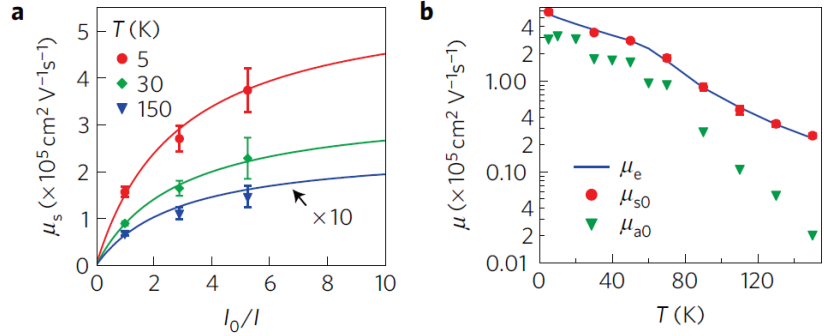

**Figure 5 | Spin mobility. a**, Spin mobility $\mu_s$ as a function of inverse laser intensity $I_0/I$, where $I_0 = 0.25$ µJ/cm², at various $T$. The solid lines are fits as described in the text, from which we determine $\mu_{s0}$, the spin mobility in the limit of zero laser intensity. **b**. Electron transport mobility, $\mu_e$, compared with spin and ambipolar mobility in the limit of zero laser intensity ($\mu_{s0}$ and $\mu_{a0}$ respectively). In this limit, the spin mobility is equal to the electron mobility, despite the crossover to coherent precession that takes place as the temperature is lowered. Error bars (s.d.) represent weighted uncertainty carried through several fitting steps as described in Methods.

Overall, the phenomena described above demonstrate that spin density propagates in a Fermi sea as a distinct, neutral degree of freedom that couples to electron motion through various interactions (*e.g.* SO and Coulomb). Fig. 1c illustrates the distinct nature of the spin polarization – given sufficient time, the more rapidly drifting SDW will leave the unpolarized packet of *e-h* density in its wake. In the limit that the number of photoexcited carriers is much less than the equilibrium number, we find that spin density propagates at the drift velocity of the Fermi sea. While this result has been predicted theoretically in models that don't include SO-induced precession [29,30], it is quite striking that this equality is preserved even as spin dynamics cross over from incoherent relaxation to coherent precession with decreasing $T$. Hopefully, the ability to measure spin mobility *via* the Doppler effect will encourage a deeper understanding of the underlying physics of spin propagation in SO coupled metallic systems, providing the basis for extending the temperature range of spin-based logic.

**Methods**

**Sample Preparation.** The experiments were performed on a 9nm wide *n*-doped GaAs/AlGaAs quantum well grown by molecular beam epitaxy on a semi-insulating GaAs (001) substrate (VB0355). A 200 nm $Al_{0.55}Ga_{0.45}As$ etch stop layer was grown on the substrate, followed by a second 10 nm etch stop of GaAs. The lower barrier was 210 nm $Al_{0.24}Ga_{0.76}As$ with Si δ-doping 95 nm below the 9 nm GaAs quantum well. The upper barrier of 190 nm $Al_{0.24}Ga_{0.76}$ includes a δ-doped layer 75 nm above the quantum well. The top layer is a 10 nm GaAs cap. The slight asymmetry in the δ-doped layers compensates for the upward drift of the Si atoms and results in nearly symmetric doping when the growth is complete. The 2DEG channel was defined by mesa etching, and Ohmic contact was made by annealing NiGeAu into the sample. After patterning, the sample was epoxied onto a non-birefringent sapphire disk and mechanically thinned to 50 microns. A citric acid etch removed the remaining GaAs substrate. The final thickness was reached after a second selective etch with hydrofluoric acid to remove the $Al_{0.55}Ga_{0.45}As$ layer. After processing, the carrier density and mobility of the 2DEG in the dark were $1.9 \times 10^{11}$ cm$^{-2}$ and $5.5 \times 10^5$ cm$^2$/V-s at 5 K, respectively, as measured by standard Van der Pauw techniques at Sandia National Lab. The sample resistance was also monitored during the course of the optical measurements at Lawrence Berkeley National Lab. In the temperature range of the experiments reported here, very little difference in the sample resistance was observed between light and dark cool-down conditions (See supplemental material section III).

**Error analysis.** The error bars in Fig. 5 stem from the uncertainty in fitting the Doppler velocimetry data such as that in Fig. 4a and 4d. The uncertainty in these fits (barely visible error bars in Figs. 4b and 4f) was used to weight subsequent fitting of the phase velocity $\dot{\phi}(q)$, and drift velocity $v_d$. The spin mobility $\mu_s(I,T)$ was obtained through normalization by the applied field. The uncertainty in $\mu_s(I,T)$, which is shown at several temperatures as error bars in Fig. 5a, was then used to weight the fits from which we extracted the values of $\mu_{s0}(T)$ shown in Fig 5b. The error bars shown in Fig. 5b for $\mu_{s0}(T)$ are smaller than the circles for almost all of the data points. Error bars for $\mu_{a0}(T)$ are even smaller and cannot be seen on the plot. Although the mobility decreases with $T$ between 30 and 150 K the relative uncertainty in $\mu_{s0}(T)$ remains essentially the same. This is the case because the measurements were performed while maintaining a constant current by increasing the applied voltage. In all cases the applied fields are small and well inside the linear response regime.

## ACKNOWLEDGMENTS

All the optical and some of the electrical measurements were carried out at Lawrence Berkeley National Laboratory and were supported by the Director, Office of Science, Office of Basic Energy Sciences, Materials Sciences and Engineering Division, of the U.S. Department of Energy under Contract No. DE-AC02-05CH11231. Sample growth and processing and some of the transport measurements were performed at the Center for Integrated Nanotechnologies, a U.S. Department of Energy, Office of Basic Energy Sciences user facility at Sandia National Laboratories (Contract No. DE-AC04-94AL85000).

## AUTHOR CONTRIBUTIONS

L.Y., J.D.K., J.O., and M.P.L. devised the experiment and wrote the manuscript. L.Y. and J.D.K. performed optical measurements and M.P.L. and L.Y. carried out transport measurements. L.Y. and J.O. performed analysis and theoretical modeling of the data. J.L.R. and D.R.T. carried out growth and fabrication of the quantum well device.

## ADDITIONAL INFORMATION

The authors declare no competing financial interests. Correspondence and requests for materials should be addressed to J.O.

# References

[1] Žutić, I, Fabian J. & Das Sarma, S. Spintronics: Fundamental and applications. *Rev. Mod. Phys.* **76,** 323-410 (2004).

[2] Dietl, T., Awschalom, D. D., Kaminska, M. & Ohno, H. Eds., *Spintronics*, vol. 82 of *Semiconductors and Semimetals* (Elsevier, Amsterdam, 2008).

[3] Koo, H. C. *et al.* Control of spin precession in a spin-injected field effect transistor. *Science* **325,** 1515-1518 (2009).

[4] Wunderlich, J. *et al.* Spin Hall Effect Transistor. *Science* **330,** 1801-1804 (2010).

[5] Kikkawa, J. M. & Awschalom, D.D. Lateral drag of spin coherence in gallium arsenide. *Nature* **397,** 139-141 (1999).

[6] Eichler, H. J., Gunter, P. & Pohl, D. W. Laser- Induced Dynamic Gratings (Springer-Verlag, Berlin, 1986).

[7] Cameron, A. R., Riblet, P. & Miller, A. Spin gratings and the measurement of electron drift mobility in multiple quantum well semiconductors. *Phys. Rev. Lett.* **76,** 4793-4796 (1996).

[8] Goodno, G. D., Dadusc, G. & Miller, R. J. D. Ultrafast heterodyne-detected transient-grating spectroscopy using diffractive optics *J. Opt. Soc. Am. B* **15,** 1791-1794 (1998).

[9] Maznev, A. A., Nelson K. A., & Rogers, J.A. Optical heterodyne detection of laser-induced gratings. *Opt. Lett.*, **23,** 1319-1321 (1998).

[10] Gedik, N. & Orenstein, J. Absolute phase measurement in heterodyne detection of transient gratings. *Opt. Lett.* **29,** 2109-2111 (2004).

[11] Dresselhaus, G. Spin–orbit coupling effects in zinc blende structures. Phys. Rev. **100**, 580–586 (1955).

[12] Schliemann, J., Egues, J. C. & Loss, D. Nonballistic spin-field-effect transistor. *Phys. Rev. Lett.* **90**, 146801 (2003).

[13] Burkov, A. A., Nunez, A. S. & MacDonald, A. H. Theory of spin-charge-coupled transport in a two-dimensional electron gas with Rashba spin-orbit interactions. *Phys. Rev. B* **70,** 155308 (2004).

[14] Bernevig, B. A., Orenstein, J. & Zhang, S.-C. Exact SU(2) Symmetry and persistent spin helix in a spin-orbit coupled system. *Phys. Rev. Lett.* **97**, 236601 (2006).

[15] Stanescu, T. D. & Galitski, V. Spin relaxation in a generic two-dimensional spin orbit coupled system. *Phys. Rev. B* **75**, 125307 (2007).

[16] Weber, C. P. *et al*. Nondiffusive spin dynamics in a two-dimensional electron gas. *Phys. Rev. Lett.* **98,** 076604 (2007).

[17] Koralek, J. D. *et al*. Emergence of the persistent spin helix in semiconductor quantum wells. *Nature* **458,** 610-613 (2009).

[18] D'Amico, I. & Vignale, G. Theory of spin Coulomb drag in spin-polarized transport. *Phys.Rev. B* **62,** 4853-4857 (2000).

[19] Weber, C.P. *et al*. Observation of spin-Coulomb drag in a two-dimensional electron gas. *Nature* **437,** 1330-1333 (2005).

[20] Flensberg, K., Jensen, T. S. & Mortensen, N. A. Diffusion equation and spin drag in spin-polarized transport. *Phys. Rev. B* **64,** 245308 (2001).

[21] D'Amico, I. & Vignale, G. Spin coulomb drag in the two-dimensional electron liquid. *Phys. Rev. B* **68,** 45307 (2003).

[22] Kleinert, P. & Bryksin, V.V. Spin polarization in biased Rashba-Dresselhaus two-dimensional electron systems. *Phys. Rev. B* **76,** 205326 (2007).

[23] Yang, L., Orenstein, J. & Lee, D.-H. Random walk approach to spin dynamics in a two-dimensional electron gas with spin-orbit coupling. *Phys. Rev. B* **82,** 155324 (2010).

[24] Crooker, S. A. & Smith, D. L. Imaging spin flows in semiconductors subjected to electric, magnetic, and strain fields. *Phys. Rev. Lett.* **94,** 236601 (2005).

[25] Leyland, W. J. H. *et al.* Enhanced spin-relaxation time due to electron-electron scattering in semiconductor quantum wells. *Phys. Rev. B* **75**, 165309 (2007).

[26] Lüffe, M. C., Kailasvuori, J. & Nunner, T. S. Relaxation mechanisms of the persistent spin helix. *Phys. Rev. B* **84,** 075326 (2011)

[27] Höpfel, R. A. *et al*. Electron-hole scattering in GaAs quantum wells. *Phys. Rev. B* **37**, 6941–6954 (1988).

[28] Yang, L. *et al.* Measurement of electron-hole friction in an n-doped GaAs/AlGaAs quantum well using optical transient grating spectroscopy. *Phys. Rev. Lett.* **106**, 247401 (2011).

[29] Flatte, M. E. & Byers, J. M. Spin diffusion in semiconductors. *Phys. Rev. Lett.* **84**, 4220–4223 (2000).

[30] D'Amico, I. & Vignale, G. Spin diffusion in doped semiconductors: the role of Coulomb interactions. *Europhys. Lett.* **55,** 566-572 (2001).